\pdfoutput=1
\documentclass[12pt,a4paper]{article}

\usepackage{ifthen} 
\newboolean{pdflatex}
\setboolean{pdflatex}{true} 

\newboolean{articletitles}
\setboolean{articletitles}{true} 

\newboolean{uprightparticles}
\setboolean{uprightparticles}{false} 

\newboolean{wordcount}
\setboolean{wordcount}{false} 

\def\paperauthors{LHCb collaboration} 
\def\paperasciititle{Observation of new Xi_c^0 baryons decaying to Lambda_c^+ K^-} 
\def\papertitle{Observation of new \Xicz baryons decaying to \Lc\Km} 
\def\paperkeywords{{High Energy Physics}, {LHCb}} 
\def\papercopyright{\the\year\ CERN for the benefit of the LHCb collaboration} 
\def\paperlicence{CC BY 4.0 licence}
\def\paperlicenceurl{https://creativecommons.org/licenses/by/4.0/}

\usepackage[top=1in, bottom=1.25in, left=1in, right=1in]{geometry}

\usepackage{microtype}
\usepackage{lineno}  
\usepackage{xspace} 
\usepackage{caption} 

\usepackage{graphicx}  
\usepackage{color}
\usepackage{colortbl}
\graphicspath{{./figs/}} 

\usepackage{amsmath} 
\usepackage{amssymb}
\usepackage{amsfonts}
\usepackage{upgreek} 

\newcommand*\patchAmsMathEnvironmentForLineno[1]{%
\expandafter\let\csname old#1\expandafter\endcsname\csname #1\endcsname
\expandafter\let\csname oldend#1\expandafter\endcsname\csname
end#1\endcsname
 \renewenvironment{#1}%
   {\linenomath\csname old#1\endcsname}%
   {\csname oldend#1\endcsname\endlinenomath}%
}
\newcommand*\patchBothAmsMathEnvironmentsForLineno[1]{%
  \patchAmsMathEnvironmentForLineno{#1}%
  \patchAmsMathEnvironmentForLineno{#1*}%
}
\AtBeginDocument{%
\patchBothAmsMathEnvironmentsForLineno{equation}%
\patchBothAmsMathEnvironmentsForLineno{align}%
\patchBothAmsMathEnvironmentsForLineno{flalign}%
\patchBothAmsMathEnvironmentsForLineno{alignat}%
\patchBothAmsMathEnvironmentsForLineno{gather}%
\patchBothAmsMathEnvironmentsForLineno{multline}%
\patchBothAmsMathEnvironmentsForLineno{eqnarray}%
}

\usepackage{hyperxmp}

\usepackage[pdftex,
            pdfauthor={\paperauthors},
            pdftitle={\paperasciititle},
            pdfkeywords={\paperkeywords},
            pdfcopyright={Copyright (C) \papercopyright},
            pdflicenseurl={\paperlicenceurl}]{hyperref}

\usepackage[colorinlistoftodos,textsize=scriptsize]{todonotes}

\usepackage[bottom,flushmargin,hang,multiple]{footmisc}

\usepackage[all]{hypcap} 

\usepackage{xspace} 
\usepackage{upgreek}

\def\lhcb   {\mbox{LHCb}\xspace}

\def\MagUp {\mbox{\em Mag\kern -0.05em Up}\xspace}

\ifthenelse{\boolean{uprightparticles}}%
{

 \def\Ppi         {\ensuremath{\uppi}\xspace}

 \def\PDelta      {\ensuremath{\Delta}\xspace}                 
 \def\PXi         {\ensuremath{\Xi}\xspace}                 
 \def\PLambda     {\ensuremath{\Lambda}\xspace}                 
 \def\PSigma      {\ensuremath{\Sigma}\xspace}                 
 \def\POmega      {\ensuremath{\Omega}\xspace}                 
 \def\PUpsilon    {\ensuremath{\Upsilon}\xspace}

 \def\PB      {\ensuremath{\mathrm{B}}\xspace}                 
                  
 \def\PD      {\ensuremath{\mathrm{D}}\xspace}

 \def\PK      {\ensuremath{\mathrm{K}}\xspace}

 \def\Pb      {\ensuremath{\mathrm{b}}\xspace}                 
 \def\Pc      {\ensuremath{\mathrm{c}}\xspace}

 \def\Pi      {\ensuremath{\mathrm{i}}\xspace}

 \def\Pp      {\ensuremath{\mathrm{p}}\xspace}

 \def\Ps      {\ensuremath{\mathrm{s}}\xspace}

 \def\thebaroffset{0.0em}
}
{

 \def\Ppi         {\ensuremath{\pi}\xspace}

 \mathchardef\PDelta="7101
 \mathchardef\PXi="7104
 \mathchardef\PLambda="7103
 \mathchardef\PSigma="7106
 \mathchardef\POmega="710A
 \mathchardef\PUpsilon="7107
                  
 \def\PB      {\ensuremath{B}\xspace}                 
                  
 \def\PD      {\ensuremath{D}\xspace}

 \def\PK      {\ensuremath{K}\xspace}

 \def\Pb      {\ensuremath{b}\xspace}                 
 \def\Pc      {\ensuremath{c}\xspace}

 \def\Pi      {\ensuremath{i}\xspace}

 \def\Pp      {\ensuremath{p}\xspace}

 \def\Ps      {\ensuremath{s}\xspace}

 \def\thebaroffset{0.18em}
}
\newcommand{\offsetoverline}[2][\thebaroffset]{\kern #1\overline{\kern -#1 #2}}%

\makeatletter
\ifcase \@ptsize \relax
  \newcommand{\miniscule}{\@setfontsize\miniscule{4}{5}}
\or
  \newcommand{\miniscule}{\@setfontsize\miniscule{5}{6}}
\or
  \newcommand{\miniscule}{\@setfontsize\miniscule{5}{6}}
\fi
\makeatother

\DeclareRobustCommand{\optbar}[1]{\shortstack{{\miniscule (\rule[.5ex]{1.25em}{.18mm})}
  \\ [-.7ex] $#1$}}

\def\squark    {{\ensuremath{\Ps}}\xspace}

\def\cquark    {{\ensuremath{\Pc}}\xspace}

\def\bquark    {{\ensuremath{\Pb}}\xspace}

\def\pion   {{\ensuremath{\Ppi}}\xspace}

\def\pip    {{\ensuremath{\pion^+}}\xspace}
\def\pim    {{\ensuremath{\pion^-}}\xspace}

\def\kaon    {{\ensuremath{\PK}}\xspace}
\def\Kbar    {{\ensuremath{\offsetoverline{\PK}}}\xspace}

\def\KorKbar {\kern \thebaroffset\optbar{\kern -\thebaroffset \PK}{}\xspace}

\def\Kzb     {{\ensuremath{\Kbar{}^0}}\xspace}
\def\Kp      {{\ensuremath{\kaon^+}}\xspace}
\def\Km      {{\ensuremath{\kaon^-}}\xspace}

\def\KS      {{\ensuremath{\kaon^0_{\mathrm{S}}}}\xspace}

\def\D       {{\ensuremath{\PD}}\xspace}

\def\DorDbar {\kern \thebaroffset\optbar{\kern -\thebaroffset \PD}\xspace}

\def\Dp      {{\ensuremath{\D^+}}\xspace}
\def\Dm      {{\ensuremath{\D^-}}\xspace}

\def\DpDm    {\ensuremath{\Dp {\kern -0.16em \Dm}}\xspace}

\def\Dsp     {{\ensuremath{\D^+_\squark}}\xspace}

\def\B       {{\ensuremath{\PB}}\xspace}
\def\Bbar    {{\ensuremath{\offsetoverline{\PB}}}\xspace}

\def\BorBbar {\kern \thebaroffset\optbar{\kern -\thebaroffset \PB}\xspace}

\def\Bzb     {{\ensuremath{\Bbar{}^0}}\xspace}
\def\Bd      {{\ensuremath{\B^0}}\xspace}

\def\BdorBdbar {\kern \thebaroffset\optbar{\kern -\thebaroffset \Bd}\xspace}

\def\Bub     {{\ensuremath{\B^-}}\xspace}

\def\Bm      {{\ensuremath{\Bub}}\xspace}

\def\Bs      {{\ensuremath{\B^0_\squark}}\xspace}

\def\BsorBsbar {\kern \thebaroffset\optbar{\kern -\thebaroffset \Bs}\xspace}

\def\Y#1S{\ensuremath{\PUpsilon{(#1S)}}\xspace}

\def\proton      {{\ensuremath{\Pp}}\xspace}

\def\Lz          {{\ensuremath{\PLambda}}\xspace}
\def\Lbar        {{\ensuremath{\offsetoverline{\PLambda}}}\xspace}
\def\LorLbar     {\kern \thebaroffset\optbar{\kern -\thebaroffset \PLambda}\xspace}

\def\Sigmares    {{\ensuremath{\PSigma}}\xspace}

\def\Xires       {{\ensuremath{\PXi}}\xspace}

\def\Omegares    {{\ensuremath{\POmega}}\xspace}

\def\Lc          {{\ensuremath{\Lz^+_\cquark}}\xspace}
\def\Lcbar       {{\ensuremath{\Lbar{}^-_\cquark}}\xspace}
\def\Xic         {{\ensuremath{\Xires_\cquark}}\xspace}
\def\Xicz        {{\ensuremath{\Xires^0_\cquark}}\xspace}
\def\Xicp        {{\ensuremath{\Xires^+_\cquark}}\xspace}

\def\Omegac      {{\ensuremath{\Omegares^0_\cquark}}\xspace}

\def\Xibz         {{\ensuremath{\Xires^0_\bquark}}\xspace}

\def\Omegab       {{\ensuremath{\Omegares^-_\bquark}}\xspace}

\newcommand{\decay}[2]{\ensuremath{#1\!\to #2}\xspace} 

\def\to                 {\ensuremath{\rightarrow}\xspace}

\def\AT#1     {\ensuremath{A_{\mathrm{T}}^{#1}}\xspace}           

\def\C#1      {\ensuremath{\mathcal{C}_{#1}}\xspace}                       
\def\Cp#1     {\ensuremath{\mathcal{C}_{#1}^{'}}\xspace}                    
\def\Ceff#1   {\ensuremath{\mathcal{C}_{#1}^{\mathrm{(eff)}}}\xspace}        
\def\Cpeff#1  {\ensuremath{\mathcal{C}_{#1}^{'\mathrm{(eff)}}}\xspace}       
\def\Ope#1    {\ensuremath{\mathcal{O}_{#1}}\xspace}                       
\def\Opep#1   {\ensuremath{\mathcal{O}_{#1}^{'}}\xspace}                    

\newcommand{\aunit}[1]{\ensuremath{\text{\,#1}}}       

\newcommand{\tev}{\aunit{Te\kern -0.1em V}\xspace}
\newcommand{\gev}{\aunit{Ge\kern -0.1em V}\xspace}
\newcommand{\mev}{\aunit{Me\kern -0.1em V}\xspace}
\newcommand{\kev}{\aunit{ke\kern -0.1em V}\xspace}
\newcommand{\ev}{\aunit{e\kern -0.1em V}\xspace}
 
\newcommand{\mevc}{\ensuremath{\aunit{Me\kern -0.1em V\!/}c}\xspace}
\newcommand{\gevc}{\ensuremath{\aunit{Ge\kern -0.1em V\!/}c}\xspace}
\newcommand{\mevcc}{\ensuremath{\aunit{Me\kern -0.1em V\!/}c^2}\xspace}
\newcommand{\gevcc}{\ensuremath{\aunit{Ge\kern -0.1em V\!/}c^2}\xspace}

\def\fb   {\ensuremath{\aunit{fb}}\xspace}
\def\invfb   {\ensuremath{\fb^{-1}}\xspace}

\def\ps   {\ensuremath{\aunit{ps}}\xspace}

\newcommand{\chisq}{\ensuremath{\chi^2}\xspace}

\newcommand{\chisqip}{\ensuremath{\chi^2_{\text{IP}}}\xspace}

\def\gsim{{~\raise.15em\hbox{$>$}\kern-.85em
          \lower.35em\hbox{$\sim$}~}\xspace}
\def\lsim{{~\raise.15em\hbox{$<$}\kern-.85em
          \lower.35em\hbox{$\sim$}~}\xspace}

\def\pt         {\ensuremath{p_{\mathrm{T}}}\xspace}

\def\tell1  {TELL1\xspace}
\def\ukl1   {UKL1\xspace}

\newcommand{\Xicstst}[1]{\ensuremath{\Xic(#1)^0}\xspace} 

\def\Xiczst      {{\ensuremath{\Xires^{**0}_\cquark}}\xspace}

\def\Xicprime        {{\ensuremath{\Xires^\prime_\cquark}}\xspace}

\def\Omegacbis     {{\ensuremath{\Omegares_\cquark}}\xspace}

\def\Sigmac       {{\ensuremath{\Sigmares_\cquark}}\xspace}

\usepackage{cite} 
\usepackage{mciteplus}

\usepackage{longtable} 

\begin{document}

\renewcommand{\thefootnote}{\fnsymbol{footnote}}
\setcounter{footnote}{1}

\ifthenelse{\boolean{wordcount}}{}{

\begin{titlepage}
\pagenumbering{roman}

\vspace*{-1.5cm}
\centerline{\large EUROPEAN ORGANIZATION FOR NUCLEAR RESEARCH (CERN)}
\vspace*{1.5cm}
\noindent
\begin{tabular*}{\linewidth}{lc@{\extracolsep{\fill}}r@{\extracolsep{0pt}}}
\ifthenelse{\boolean{pdflatex}}
{\vspace*{-1.5cm}\mbox{\!\!\!\includegraphics[width=.14\textwidth]{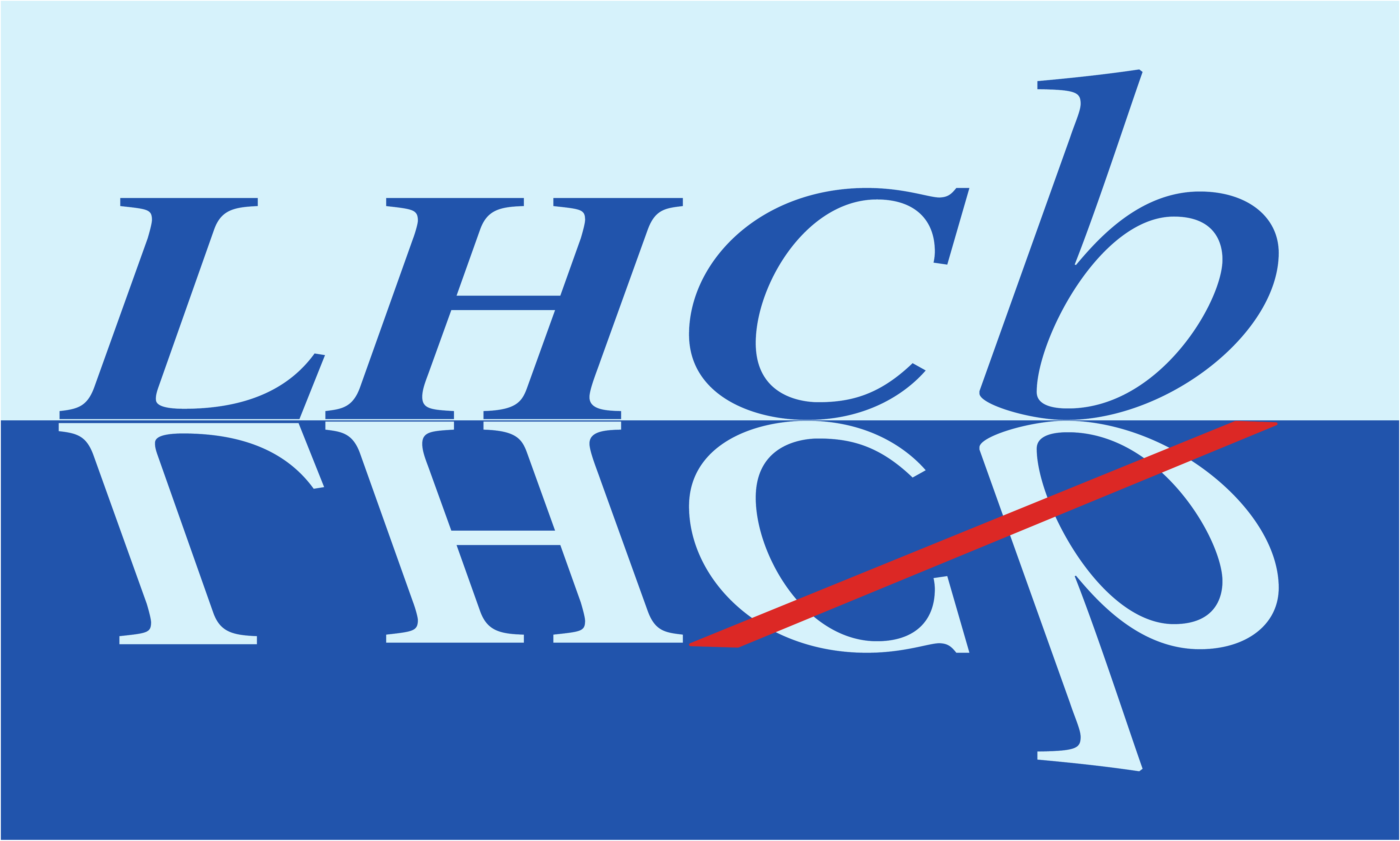}} & &}%
{\vspace*{-1.2cm}\mbox{\!\!\!\includegraphics[width=.12\textwidth]{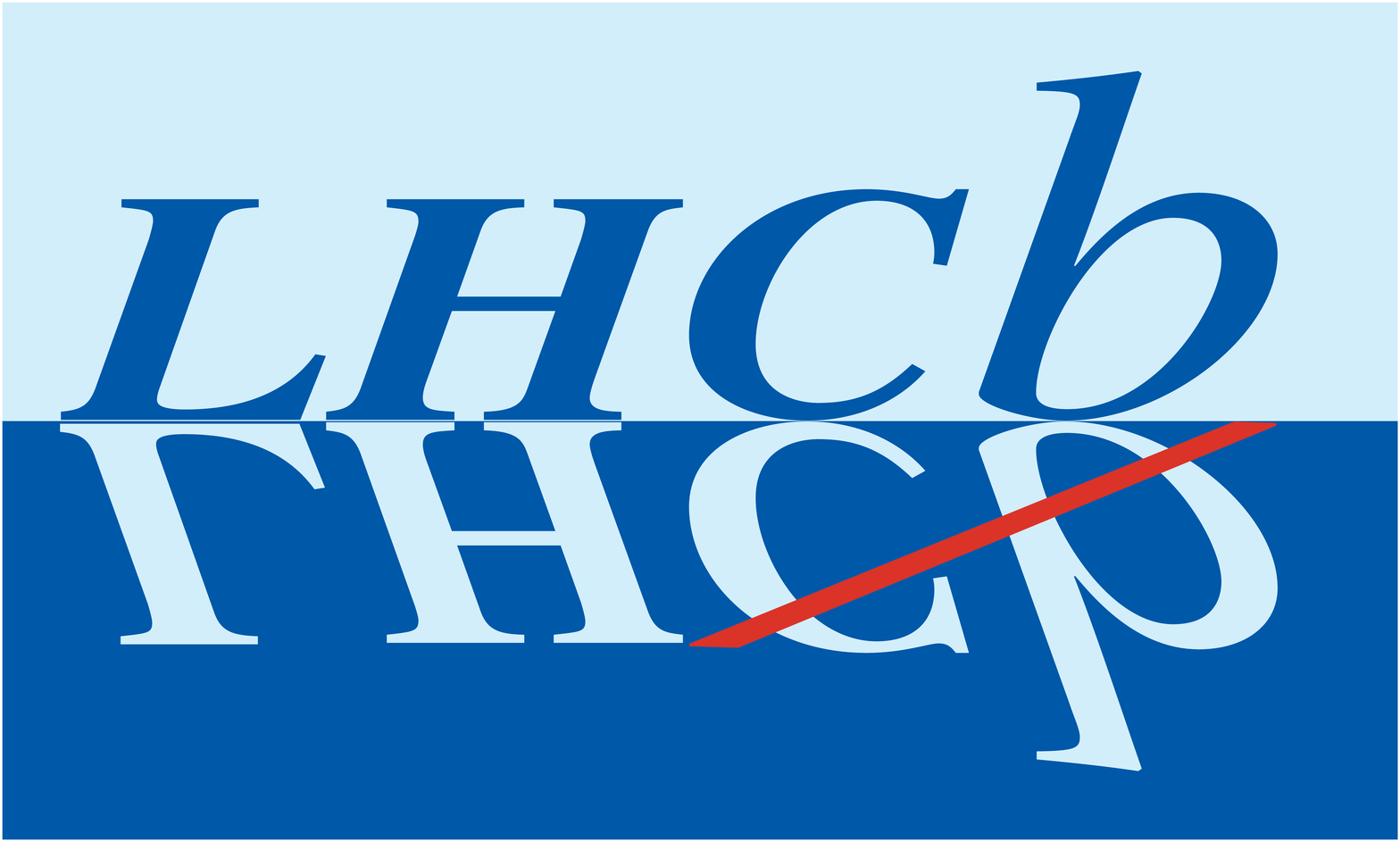}} & &}%
\\
 & & CERN-EP-2020-038 \\  
 & & LHCb-PAPER-2020-004 \\  
 & & 04 June 2020 \\ 
 & & \\
\end{tabular*}

\vspace*{2.cm}

{\normalfont\bfseries\boldmath\huge
\begin{center}
  \papertitle 
\end{center}
}

\vspace*{1.0cm}

\begin{center}
\paperauthors\footnote{Authors are listed at the end of this paper.}
\end{center}

\vspace{\fill}

\begin{abstract}
  \noindent
  The \Lc\Km mass spectrum is studied with a data sample of \proton\proton collisions at a centre-of-mass energy of 13\tev corresponding to an integrated luminosity of 5.6\invfb collected by the \lhcb experiment. Three \Xicz states are observed with a large significance and their masses and natural widths are measured to be 
  \begin{eqnarray*}
  m(\Xicstst{2923})&=& 2923.04 \pm 0.25 \pm 0.20 \pm 0.14\mev,\\
  \Gamma(\Xicstst{2923}) &=& 7.1 \pm 0.8 \pm 1.8\mev,
  \end{eqnarray*}\vspace{-0.7cm}
  \begin{eqnarray*}
  m(\Xicstst{2939}) &=& 2938.55 \pm 0.21 \pm 0.17 \pm 0.14\mev,\\
  \Gamma(\Xicstst{2939}) &=& 10.2 \pm 0.8 \pm 1.1\mev,
   \end{eqnarray*}\vspace{-0.7cm}
  \begin{eqnarray*}
  m(\Xicstst{2965}) &=& 2964.88 \pm 0.26 \pm 0.14 \pm 0.14\mev, \\
  \Gamma(\Xicstst{2965}) &=& 14.1 \pm 0.9 \pm 1.3\mev,
  \end{eqnarray*}
  where the uncertainties are statistical, systematic, and due to the limited knowledge of the \Lc mass. The \Xicstst{2923} and \Xicstst{2939} baryons are new states. The $\Xicstst{2965}$ state is in the vicinity of the known $\Xicstst{2970}$ baryon; however, their masses and natural widths differ significantly.

  
\end{abstract}

\vspace*{2.0cm}

\begin{center}
  Published in Phys. Rev. Lett. 124 (2020) 222001
\end{center}

\vspace{\fill}

{\footnotesize 
\centerline{\copyright~\papercopyright. \href{\paperlicenceurl}{\paperlicence}.}}
\vspace*{2mm}

\end{titlepage}


\newpage
\setcounter{page}{2}
\mbox{~}
%
%
%
%

}

\renewcommand{\thefootnote}{\arabic{footnote}}
\setcounter{footnote}{0}

\cleardoublepage


\pagestyle{plain} 
\setcounter{page}{1}
\pagenumbering{arabic}


%

Singly charmed baryons are composed of a charm quark and two light quarks. 
Due to the large mass difference between the charm and the lighter quarks, these 
baryons provide an insight into the spectrum of states using symmetries described by the Heavy
Quark Effective Theory~\cite{Grozin:1992yq,Mannel:1996rg}. Numerous theoretical predictions of the properties of heavy baryons, containing either a charm or a beauty quark, have been
made in recent years~\cite{Ebert:2007nw, Roberts:2007ni, Garcilazo:2007eh, Migura:2006ep, Ebert:2011kk, Valcarce:2008dr, Shah:2016nxi, Vijande:2012mk, Yoshida:2015tia, Chen:2015kpa, Chen:2016phw}. In many of these models, the heavy quark interacts with
a lighter diquark, which is treated as a single object. Other predictions are based on
Lattice QCD calculations~\cite{Padmanath:2013bla}.

In $2017$, the \lhcb collaboration reported the observation of five new narrow \Omegac baryons decaying 
to the \Xicp\Km final state~\cite{LHCb-PAPER-2017-002}, four of which were later confirmed by the Belle collaboration~\cite{Yelton:2017qxg}. It is currently not understood why the natural widths of these resonances
are small~\cite{Chiladze:1997ev,Karliner:2017kfm}, although a similar trend has recently been observed in the excited \Omegab states decaying to \Xibz\Km~\cite{LHCb-PAPER-2019-042}.
Investigating a different charmed mass spectrum 
could lead to a better understanding of this feature.

A natural extension to the \Xicp\Km analysis is the study of the \Lc\Km spectrum. 
The BaBar collaboration was the first to observe a structure in the \Lc\Km mass
spectrum in \decay{\Bm}{\Km\Lc\Lcbar} decays peaking at 2.93\gev in 2007~\cite{Aubert:2007eb}. However, it was not interpreted as
a new state due to the absence of an amplitude analysis. Unless otherwise stated, charge-conjugate processes are implicitly included, and natural units with $\hbar=c=1$ are 
used throughout. Later that year another analysis was published~\cite{Aubert:2007dt}, looking at strongly interacting prompt decays of charm-strange baryons to several final states,
one of which was \Lc\Km. No resonances were reported in the \Lc\Km mass spectrum. 
The Belle collaboration also reported the study of \decay{\Bm}{\Km\Lc\Lcbar} decays~\cite{Li:2017uvv}. A peaking structure
was observed in the \Lc\Km mass spectrum
compatible with the results of Ref.~\cite{Aubert:2007eb}
and interpreted as 
a new \Xicz baryon, dubbed $\Xicstst{2930}$. Similarly, evidence of the isospin partner $\Xic(2930)^+$ in \decay{\Bzb}{\Kzb\Lc\Lcbar}
decays has been claimed~\cite{Li:2018fmq}.

This letter presents a search for excited \Xicz baryons, hereafter referred to as \Xiczst, in the \Lc\Km spectrum in a mass region around the  \Xicstst{2930} state, with the \Lc baryons reconstructed in the \proton\Km\pip final state. Defining $\Delta M \equiv m(\Lc\Km) - m(\Lc) - m(\Km)$, the region considered is $\Delta M <300\mev$.
The data are collected in \proton\proton collisions with the \lhcb detector at a
centre-of-mass energy of 13\tev, corresponding to an integrated luminosity of 5.6\invfb. 

The \lhcb detector~\cite{LHCb-DP-2008-001,LHCb-DP-2014-002} is a single-arm forward spectrometer covering the pseudorapidity range $2 < \eta < 5$, designed for the study of particles containing \bquark\ or \cquark\ quarks. The detector elements that are particularly relevant to this analysis are: a silicon-strip vertex detector surrounding the $\proton\proton$ interaction region that allows \cquark\ and \bquark\ hadrons to be identified from their characteristically long flight distance; a tracking system that provides a measurement of the momentum of charged particles; and two ring-imaging Cherenkov detectors that are able to discriminate between different species of charged hadrons.
The online event selection 
is performed by a trigger, which consists of a hardware stage, based on information from the 
calorimeter and muon systems, followed by a two-level software stage, which applies a full 
event reconstruction\cite{LHCb-DP-2012-004,LHCb-DP-2016-001}. Simulated data samples are produced 
with the software packages described in
Refs.~\cite{Sjostrand:2007gs,LHCb-PROC-2010-056,Lange:2001uf,Allison:2006ve,LHCb-PROC-2011-006} and are used to optimise the selection requirements, to quantify the 
invariant-mass resolution, and to model physics processes which may constitute peaking backgrounds in the analysis.

Candidate \Lc baryons are formed from the combination of three tracks of good quality which are inconsistent with originating from any primary proton-proton interaction vertex~(PV) and have large transverse momentum~(\pt).  Particle identification (PID) requirements are imposed on all three tracks to suppress combinatorial background and misidentified charm-meson decays.
The \Lc candidates are required to have $\pt > 2\gev$ and are constrained to originate from the associated PV by requiring a small \chisqip,  defined as the difference between the vertex-fit \chisq of the PV reconstructed with and without the candidate in question. 
The \Lc vertex must also be displaced from the associated PV such that the \Lc decay time is longer than 0.3\ps.
A multivariate classifier based on a
boosted decision tree~(BDT) algorithm~\cite{Breiman,AdaBoost} implemented in the TMVA toolkit~\cite{Hocker:2007ht,*TMVA4} is used to further improve the \Lc signal purity. 
The input variables given to the BDT are the $\chi^2$ value of the \Lc  decay-vertex fit, the \Lc flight distance between the production and decay vertex, the angle between the \Lc momentum vector and the line that 
joins the \Lc decay vertex with its associated PV, the \chisqip and \pt of the \Lc candidate, and the \chisqip and PID responses of the \Lc decay particles. 
The background 
sample used in the BDT training  consists of the lower and upper
sidebands of the \proton\Km\pip invariant mass distribution,
$2230-2250\mev$ and $2320-2340\mev$,  respectively. The signal sample used is the \Lc sample in data after subtracting the background by means of the $sPlot$ technique\cite{Pivk:2004ty}, exploiting $m(\proton\Km\pip)$ as discriminating variable. 
The training of the multivariate algorithm is carried out by using 20\,000 candidates of  the reconstructed \Lc candidates from the  data recorded in 2016.
The requirement on the BDT response is determined using 200\,000 \Lc candidates 
by maximising the figure of merit $S/\sqrt{S+B}$, where $S$ is the \Lc signal yield extracted from a fit to the mass spectrum of \Lc candidates passing a given BDT requirement and $B$ is expected background yield. The value for $B$ is extrapolated by scaling the background yield over the full mass range of the fit to a $\pm 15\mev$ mass range around the \Lc peak.

\begin{figure}[tb]
  \begin{center}
  \ifthenelse{\boolean{wordcount}}{}{
    \includegraphics[width=0.49\linewidth]{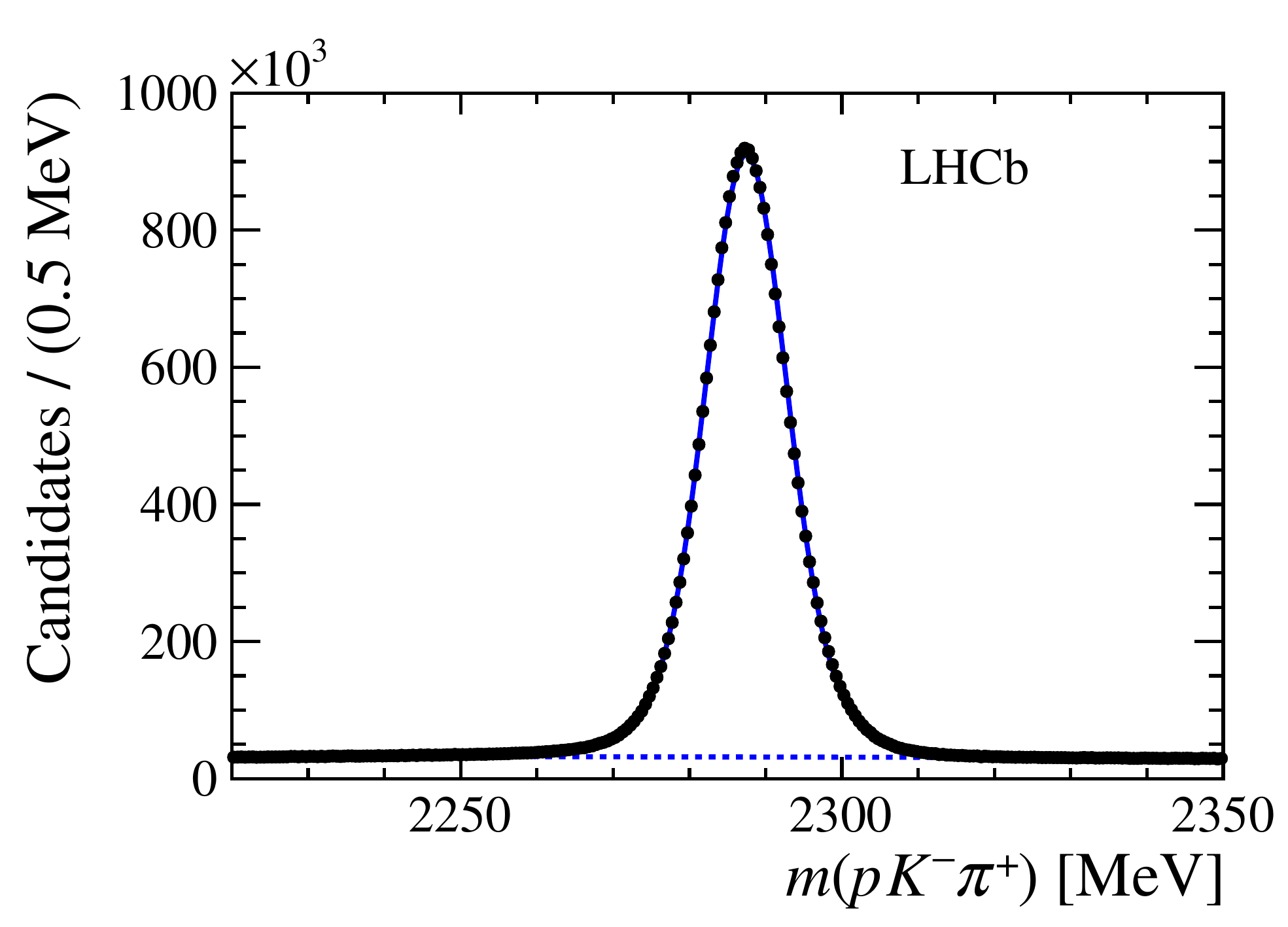}
    \vspace*{-0.5cm}}
  \end{center}
  \caption{
    Distribution of the reconstructed invariant mass $m(\proton\Km\pip)$ for 20$\%$ of the candidates in the \Lc 
    sample passing the selection described in the text. The solid blue curve shows the result of the fit, 
    and the dashed blue line indicates the background component of the fit.}
  \label{fig:Lc}
\end{figure}

Misidentified \decay{\Dp}{\Km\pip\pip}, \decay{\Dp}{\Kp\Km\pip} and \decay{\Dsp}{\Kp\Km\pip} background decays are observed after changing the mass hypothesis of the proton into a kaon or a pion.
These background components are reduced by employing a tighter PID selection and requiring the invariant mass $m(\Kp\Km)$ to differ by at least 10\mev from the known $\phi(1020)$ mass~\cite{PDG2019}. 
Removing all candidates in mass windows around the $\PD_{(\squark)}^+$ mass distributions
would result in a large loss of signal efficiency and therefore is not implemented. However, it is checked that the results of the analysis are stable when these background components are removed fully. 
About 125 million \Lc signal decays are selected for further analysis with a purity of 93\%. The invariant-mass distribution of 20\% of the \Lc candidates satisfying these selection requirements is shown in
Fig.~\ref{fig:Lc}. 

The \Xiczst candidates are formed from \Lc\Km combinations, where the \Lc candidate mass is required to 
be within 20\mev of the known \Lc mass~\cite{PDG2019}.
Each \Lc candidate is combined with a \Km 
candidate that is consistent with originating from the associated PV. The \Lc and \Km particles are 
fitted to a common vertex, which is required to be consistent with the associated PV.

The main contribution to the combinatorial background in the \Lc\Km mass spectrum is due to the large number of kaon candidates from the PV. 
The signal to background ratio is improved by optimising the PID criteria of the \Km candidates and the \pt requirement on the \Xiczst candidates using the figure of merit  $\epsilon/(\sqrt{B_P}+5/2)$~\cite{Punzi:2003bu}. 
Here, $\epsilon$ is the efficiency determined using simulated \decay{\Xicstst{2930}}{\Lc\Km} decays, and $B_P$ is the number of \Lc\Km candidates in the mass region $260 < \Delta M < 290\mev$,
corresponding to the background expected in a mass window around the expected \Xicstst{2930} signal, with width $\Gamma(\Xicstst{2930})= 26 \pm 8\mev$~\cite{PDG2019}. 
Based on the optimisation above, the \pt of the \Xicz candidates is required to be larger than 7350\mev, and the kaon PID is required to satisfy a tight criterion. The fraction of events with multiple candidates is found to be $0.88\%$ in the entire $\Delta M$ range. All candidates are included in the analysis.

\begin{figure}[!tb]
  \begin{center}
  \ifthenelse{\boolean{wordcount}}{}{
    \includegraphics[width=0.37\linewidth]{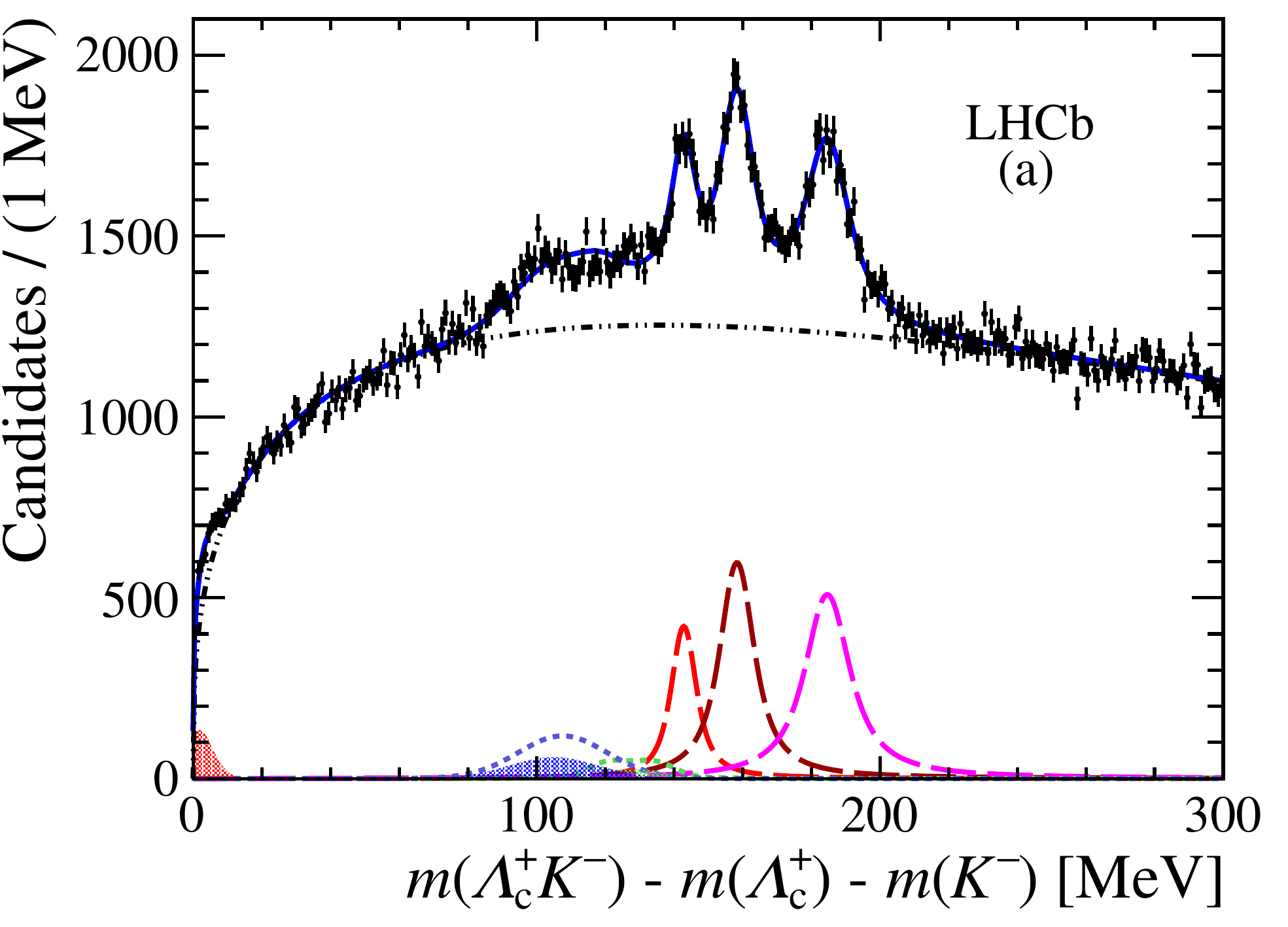}
    \includegraphics[width=0.37\linewidth]{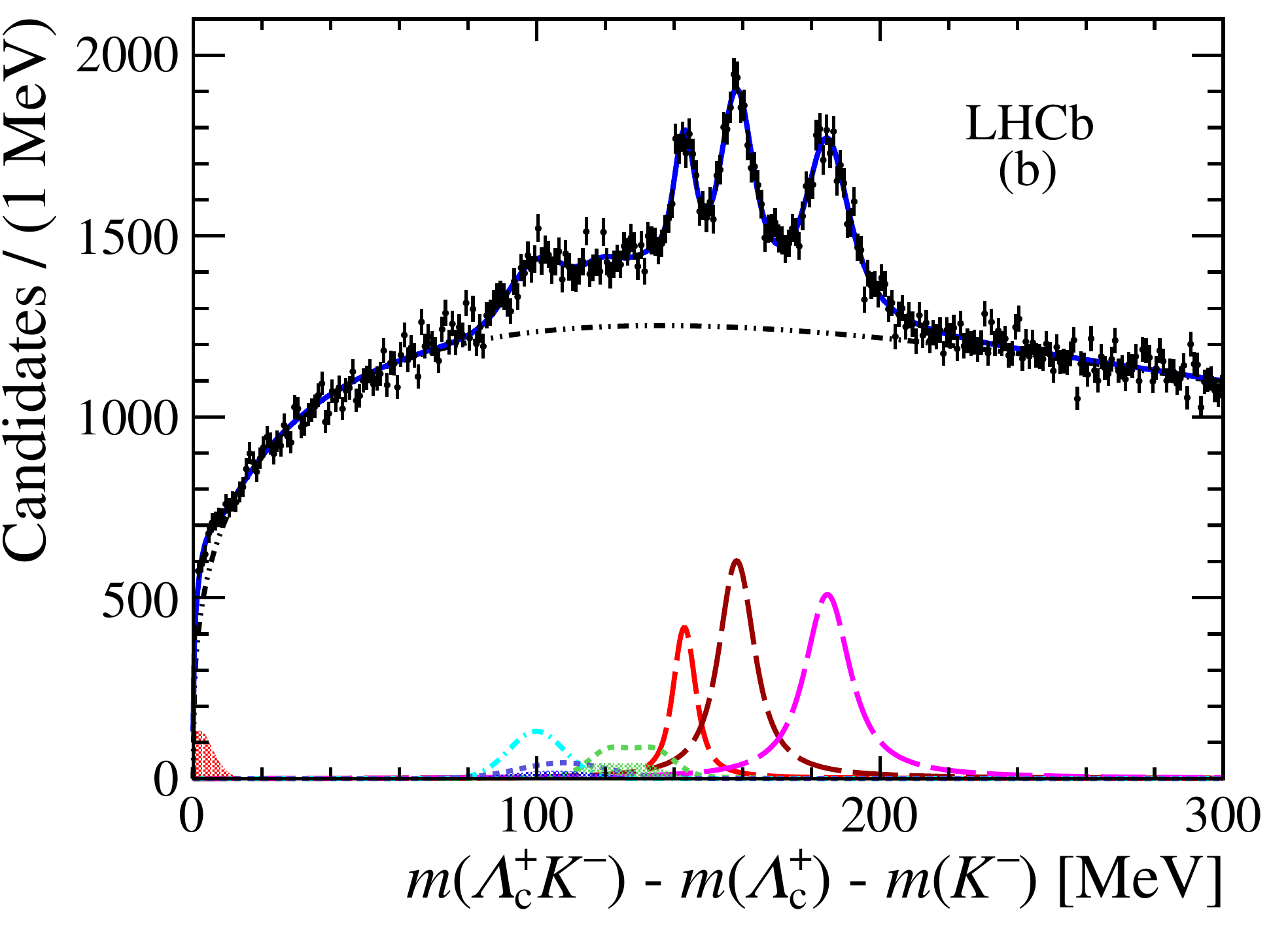}
    \includegraphics[width=0.26\linewidth]{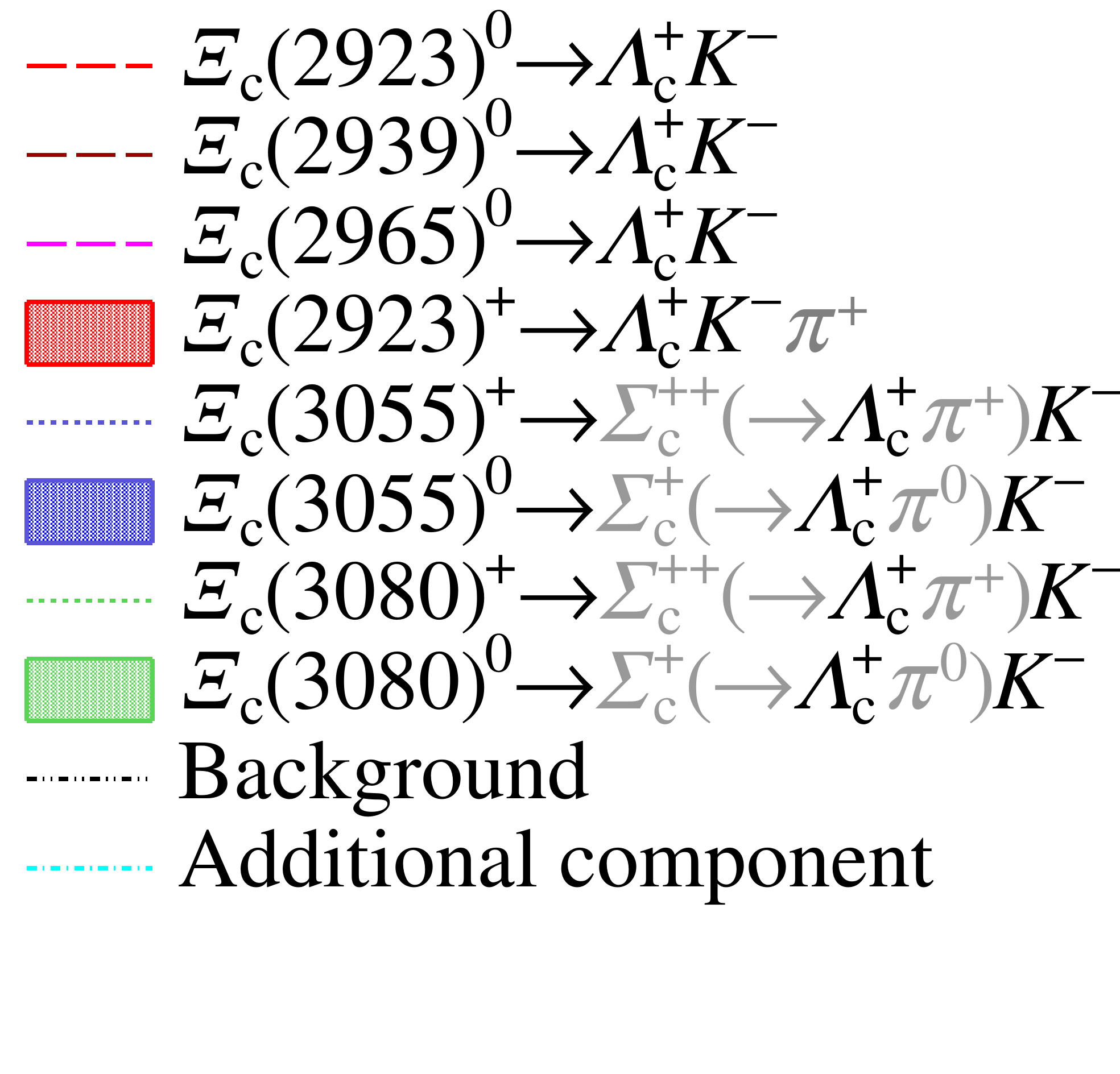}
    \vspace*{-0.5cm}}
  \end{center}
  \caption{
    Distributions of the reconstructed invariant-mass difference \mbox{$\Delta M = m(\Lc\Km) - m(\Lc) - m(\Km)$} for all candidates passing the selection
    requirements described in the text. The black symbols show the selected signal candidates. The result of a fit, described in the text, is overlaid (solid blue line). In plot (a) the reference fit is shown. Plot (b) shows an alternative description to the data, where an additional Gaussian component given by the cyan dot-dashed line is added to the fit model around $\Delta M \simeq100$\mev. The missing child particles in the reconstruction are indicated in grey in the legend.
    }
  \label{fig:LcK}
\end{figure}

The resulting $\Delta M$ distribution of the signal candidates
is shown in Fig.~\ref{fig:LcK}, where a fit to the data is superimposed. Three narrow structures are observed in the \Lc\Km candidate spectrum. These peaking structures are not seen in the
wrong-sign (WS) \Lc\Kp candidates or \Lc sideband distributions. 
The $\Delta M$ distribution also shows a broad structure to the left of the three narrow structures consistent with being partially reconstructed
\mbox{\decay{\Xic(3055)}{\decay{{{\ensuremath{\Sigmares_\cquark(2455)}}\xspace}(}{ \Lc\pi})\Km}} and \mbox{\decay{\Xic(3080)}{\decay{{{\ensuremath{\Sigmares_\cquark(2455)}}\xspace}(}{ \Lc\pi})\Km}}
decays, where the pion is not reconstructed.

An unbinned maximum-likelihood fit, henceforth denoted the reference fit, is performed to the $\Delta M$ distribution to measure the parameters of each peak. The background is modelled by an empirical function of the form $\Delta M^{a}\times \exp(-b \times \Delta M)$, where $a$ and $b$ vary freely. Each signal peak is described by an $S$-wave relativistic Breit$-$Wigner function convolved with a mass-resolution function. The experimental mass resolution is determined using simulated \decay{\Xiczst}{\Lc\Km} decays at several \Xiczst masses.
In the $\Delta M$ interval where the three narrow peaks occur, the mass resolution varies between $1.7$ and $2.2$\mev. 
Simulated data are also generated to determine the shape of partially reconstructed $\Xic(3055)$ and $\Xic(3080)$ decays.
The shapes of these contributions are allowed to shift in $\Delta M$ by the uncertainties in the decay-product masses, where the shift is Gaussian constrained. From isospin symmetry, the yields of 
the components are constrained 
to be twice as large as
the corresponding $\Xires_\cquark(3055)^0$ and $\Xires_\cquark(3080)^0$ components.
The fit model outlined so far does not accurately describe the data in the mass region close to the kinematic threshold, and thus an additional component is considered.
There are no known decays of 
\decay{{{\ensuremath{\Sigmares_\cquark(2455)}}\xspace}(}{ \Lc\pi)\Km} or
\decay{{{\ensuremath{\Sigmares_\cquark(2520)}}\xspace}(}{ \Lc\pi)\Km} 
which could enter the sample as partially reconstructed components at $\Delta M \simeq0$. It is observed that the missing component is consistent with being due to the partial reconstruction of the state that peaks around $\Delta M\simeq140$\mev when it decays directly to the \Lc\Km\pip final state without any intermediate resonance. 
The shape of these partially reconstructed decays is taken from simulated samples generated using the RapidSim package~\cite{Cowan:2016tnm} and the yield is a free parameter in the fit. 

The $\Delta M$ distribution with the fit to the data superimposed is shown in Fig.~\ref{fig:LcK}(a). The goodness-of-fit value is $\chisq/\rm{ndof}=301/(300-19) = 1.07$, 
where ndof is the number of the degrees of freedom. Table~\ref{tab:fit} shows the results for the parameters of the signal peaks of the reference fit, hereafter named \Xicstst{2923}, \Xicstst{2939} and \Xicstst{2965}. 

\begin{table}[t]
  \caption{
    \small Peak positions in the invariant-mass difference distribution $\Delta M$, natural widths $\Gamma$, signal yields and local significances of the three mass peaks obtained from the fit to the \Lc\Km mass spectrum, where the systematic uncertainties are statistical.}
    \ifthenelse{\boolean{wordcount}}{}{
\begin{center}\begin{tabular}{ccc}
    \hline
    Peak of $\Delta M$ $[\!\mev]$     & $\Gamma$  $[\!\mev]$           & Signal yields  \\ 
    \hline
    $142.91 \pm 0.25$ & $\phantom{0}7.1 \pm 0.8 $    & $\phantom{0}5400 \pm 400$ \\
    $158.45 \pm 0.21$ & $10.2 \pm 0.8 $   & $10400 \pm 600$ \\
    $184.75 \pm 0.26$ & $14.1 \pm 0.9 $   & $11700 \pm 600$ \\
    \hline
  \end{tabular}\end{center}}
\label{tab:fit}
\end{table}

To validate the presence of the signal components and test the stability of the fit parameters, several additional checks are performed. The data are fitted in samples according to the year of data-taking and to different data-taking conditions depending on the \lhcb magnet configuration. The \Lc\Km sample and its charge conjugate are also studied separately. The results are consistent among all samples.

The data and the reference fit show the least compatibility in the region around $\Delta M\simeq100$\mev. This may be due to a mismodelling of the partially reconstructed distributions, but it could also be due to the presence of further new $\Xiczst$ baryon states. 
Figure~\ref{fig:LcK}(b) shows the $\Delta M$ distribution for the signal sample where an additional component, parametrized by an empirical Gaussian function, has been added to the reference fit. This fit has a goodness-of-fit value of $\chisq/\rm{ndof}=278/(300-22) = 1.00$. As a cross-check, this structure is tested in subsamples of the data set divided by data-taking year, and showed an inconsistency in the scaling of the yield with respect to the integrated luminosity. Furthermore, the feed-down components are highly suppressed when this contribution is included. More data are required to understand the cause of this additional structure. It is accounted for when calculating the systematic uncertainties.

 \begin{table}[!t]
\centering
  \caption{\small Summary of the contributions to the systematic uncertainties on the resonance parameters. Absolute deviations from the nominal fit are quoted.}
\ifthenelse{\boolean{wordcount}}{}{
\resizebox{\columnwidth}{!}{%
  \begin{tabular}{lcccccc} \hline
Source                      & \multicolumn{2}{c}{\Xicstst{2923}} & \multicolumn{2}{c}{\Xicstst{2939}}  & \multicolumn{2}{c}{\Xicstst{2965}} \cr
                            & $m\,[\!\mev]$ & $\Gamma\,[\!\mev]$  & $m\,[\!\mev]$ & $\Gamma\,[\!\mev]$  & $m\,[\!\mev]$ & $\Gamma\,[\!\mev]$ \cr \hline
Alternative fit model       & 0.15 &  1.6           & 0.14 & 0.4          & 0.04 & 1.1 \cr
Resonance interferences     & 0.08 & 0.7          & 0.06 & 1.0          & 0.11 & 0.7 \cr
Momentum-scale & 0.04 & \phantom{0.}--\phantom{0}          & 0.05 & \phantom{0.}--\phantom{0}          & 0.06  & \phantom{0.}--\phantom{0} \cr
Energy losses               & 0.04 & \phantom{0.}--\phantom{0}                       & 0.04 & \phantom{0.}--\phantom{0}          & 0.04 & \phantom{0.}--\phantom{0} \cr
Resolution calibration         & \phantom{0.}--\phantom{0} & 0.6          & \phantom{0.}--\phantom{0} & 0.2          & \phantom{0.}--\phantom{0} & 0.3 \cr \hline
Total                       & 0.20 & 1.8          & 0.17 & 1.1          & 0.14 & 1.3 \cr \hline
  \end{tabular}
  }}
  \label{tab:syst}
 \end{table}

Several sources of systematic uncertainty may affect the measured parameters.
The fit model uncertainty is evaluated by replacing the background model by an alternative function, consisting of a combination of the wrong-sign $m(\Lc\Kp)$ invariant-mass distribution shape and the shape obtained from candidates in the \Lc sideband. In addition, the choice of the relativistic Breit$-$Wigner model is changed by setting the values of the angular momentum $L$ between the child particles to $L=1,2$ and separately varying the Blatt$-$Weisskopf factors~\cite{Blatt:1952ije} from 2 to 4 $\gev^{-1}$. Furthermore, the fit is adapted to include any partially reconstructed decays  \decay{\Xires_\cquark^{**}}{\decay{{{\ensuremath{\Sigmares_\cquark(2455/2520)}}\xspace}(}{ \Lc\pi})\Km}
that are found to not contribute significantly to the reference fit. Finally, deviations in fit parameters between the reference fit and the fit shown in Fig.~\ref{fig:LcK}(b) are included in the fit model uncertainty. The largest deviation from the reference fit is quoted as the systematic uncertainty for the fit model. 
Resonances with the same spin-parity that are close in mass can interfere. An interference term is introduced between neighbouring resonances, for one pair of resonances at a time. With the interference term, the lineshape takes the form $A = |c_j\mathrm{BW}_j+ c_k\mathrm{BW}_ke^{i\phi}|^2$ where $j$ and $k$ denote the two resonances, $\mathrm{BW}_{j,k}$ are Breit$-$Wigner functions and $c_{j,k}$ and $\phi$ are free real parameters. The largest difference between the reference fit and a fit where resonance interference is allowed is used as the systematic uncertainty.
In addition, several other sources of systematic uncertainty only affect the mass measurement. These include the momentum-scale uncertainty, evaluated by shifting the momentum-scale of charged tracks by $\pm 0.03\%$~\cite{LHCb-PAPER-2013-011} in simulated decays, and the imperfect modelling of the energy loss in the detector material, resulting in a systematic uncertainty of 0.04\mev~\cite{LHCb-PAPER-2010-001}. 
Finally, a systematic uncertainty is attributed to the width measurement, to account for the fact that the simulation may not reproduce the absolute mass resolution perfectly. 
The corresponding systematic uncertainty is obtained by the change in the width when the value of the resolution, determined on simulated data, is varied by
$10\%$~\cite{LHCb-PAPER-2014-048}. The systematic uncertainties are summarised in Table~\ref{tab:syst} and in Table~\ref{tab:results} their measured masses and natural widths are summarised.

The observations described in this letter and the lack of any \Xicstst{2930} signal indicates that the broad bump observed in  \decay{\Bm}{\Km\Lc\Lcbar} decays~\cite{Aubert:2007eb, Li:2017uvv} might be due to the overlap of two narrower states, such as  the \Xicstst{2923} and \Xicstst{2939} baryons.
The \Xicstst{2965} baryon is in the vicinity of the known \Xicstst{2970} baryon, which has been observed in different decay modes, $\Sigmac(2455)^0\KS$~\cite{Aubert:2007dt}, $\Xicprime^+\pim$~\cite{Yelton:2016fqw} and $\Xic(2645)^+\pim$~\cite{Lesiak:2008wz}.  
Furthermore, the \Xicstst{2965} resonance has a natural width and mass which differs significantly from that of the $\Xicstst{2970}$ baryon, \mbox{$\Gamma(\Xicstst{2970}) = 28.1^{+3.4}_{-4.0}\mev$} and \mbox{$m(\Xicstst{2970}) = 2967.8^{+0.9}_{-0.7}\mev$}~\cite{PDG2019}. 
Further studies are required to establish whether the \Xicstst{2965} state is indeed a different baryon. 
The equal spacing rule~\cite{GellMann:1962xb,Okubo:1961jc} succeeded to predict the mass of the \Omegares baryon and holds for other flavour multiplets such as the sextet of the $J^P=3/2^+$ charmed ground states:
\ifthenelse{\boolean{wordcount}}{}{
\begin{equation*}
m(\Omegacbis(2770)^0) - m(\Xicstst{2645}) \simeq m(\Xicstst{2645}) - m(\Sigmac(2520)^0) \simeq 125\mev.\end{equation*}}
It is noted that the rule also seems to hold for the \Xicstst{2923}, \Xicstst{2939} and \Xicstst{2965} baryons within a precision of a few\,\mev:
\ifthenelse{\boolean{wordcount}}{}{
\begin{align*}m(\Omegacbis(3050)^0) - m(\Xicstst{2923}) \simeq m(\Xicstst{2923}) - m(\Sigmac(2800)^0) &\simeq 125\mev,\\
m(\Omegacbis(3065)^0) - m(\Xicstst{2939}) &\simeq 125\mev,\\
m(\Omegacbis(3090)^0) - m(\Xicstst{2965}) &\simeq 125\mev.\end{align*}}
This pattern may indicate that the new states reported in this analysis are related to the excited \Omegac baryons observed in the \Xicp \Km spectrum. Measurements of spin-parities will be crucial to confirm whether they belong to the same flavour multiplets.

In summary, $\proton\proton$ collision data collected by the \lhcb experiment at a centre-of-mass energy of 13\tev, corresponding to an integrated luminosity of 5.6\invfb, are used to search for excited \Xicz resonances in the \Lc\Km mass spectrum. Three different \Xicz baryons, \Xicstst{2923}, \Xicstst{2939} and \Xicstst{2965}, are unambiguously observed. The two baryons at lower mass are observed for the first time, while an investigation of additional final states is required to establish whether the \Xicstst{2965} and \Xicstst{2970} states are different baryons.

\begin{table}[t]
  \caption{
    \small Summary of the parameters for the studied states, showing the measured $\Delta M$ values, the masses and the natural widths, where the first uncertainty is statistical and the second uncertainty is systematic. For the mass measurement, the third uncertainty denotes the uncertainty on the known \Lc mass~\cite{PDG2019}.}
    \ifthenelse{\boolean{wordcount}}{}{
\begin{center}\begin{tabular}{cccc}
    \hline
    Resonance & Peak of $\Delta M$ $[\!\mev]$     & Mass $[\!\mev]$ & $\Gamma$  $[\!\mev]$  \\ 
    \hline
    $\Xic(2923)^0$ & $142.91 \pm 0.25 \pm 0.20$ & $2923.04 \pm 0.25 \pm 0.20 \pm 0.14$ & $\phantom{0}7.1 \pm 0.8 \pm 1.8$   \\
    $\Xic(2939)^0$ & $158.45 \pm 0.21 \pm 0.17$ & $2938.55 \pm 0.21 \pm 0.17 \pm 0.14$& $10.2 \pm 0.8 \pm 1.1$   \\
    $\Xic(2965)^0$ & $184.75 \pm 0.26 \pm 0.14$ & $2964.88 \pm 0.26 \pm 0.14 \pm 0.14$& $14.1 \pm 0.9 \pm 1.3$  \\
    \hline
  \end{tabular}\end{center}}
\label{tab:results}
\end{table}









\ifthenelse{\boolean{wordcount}}{}{
\section*{Acknowledgements}
%
%
\noindent We express our gratitude to our colleagues in the CERN
accelerator departments for the excellent performance of the LHC. We
thank the technical and administrative staff at the LHCb
institutes.
We acknowledge support from CERN and from the national agencies:
CAPES, CNPq, FAPERJ and FINEP (Brazil); 
MOST and NSFC (China); 
CNRS/IN2P3 (France); 
BMBF, DFG and MPG (Germany); 
INFN (Italy); 
NWO (Netherlands); 
MNiSW and NCN (Poland); 
MEN/IFA (Romania); 
MSHE (Russia); 
MinECo (Spain); 
SNSF and SER (Switzerland); 
NASU (Ukraine); 
STFC (United Kingdom); 
DOE NP and NSF (USA).
We acknowledge the computing resources that are provided by CERN, IN2P3
(France), KIT and DESY (Germany), INFN (Italy), SURF (Netherlands),
PIC (Spain), GridPP (United Kingdom), RRCKI and Yandex
LLC (Russia), CSCS (Switzerland), IFIN-HH (Romania), CBPF (Brazil),
PL-GRID (Poland) and OSC (USA).
We are indebted to the communities behind the multiple open-source
software packages on which we depend.
Individual groups or members have received support from
AvH Foundation (Germany);
EPLANET, Marie Sk\l{}odowska-Curie Actions and ERC (European Union);
ANR, Labex P2IO and OCEVU, and R\'{e}gion Auvergne-Rh\^{o}ne-Alpes (France);
Key Research Program of Frontier Sciences of CAS, CAS PIFI, and the Thousand Talents Program (China);
RFBR, RSF and Yandex LLC (Russia);
GVA, XuntaGal and GENCAT (Spain);
the Royal Society
and the Leverhulme Trust (United Kingdom).

}



\ifthenelse{\boolean{wordcount}}{}{
\addcontentsline{toc}{section}{References}
\bibliographystyle{LHCb}
\bibliography{main,standard,LHCb-PAPER,LHCb-CONF,LHCb-DP,LHCb-TDR}

\ifx\mcitethebibliography\mciteundefinedmacro
\PackageError{LHCb.bst}{mciteplus.sty has not been loaded}
{This bibstyle requires the use of the mciteplus package.}\fi
\providecommand{\href}[2]{#2}
\begin{mcitethebibliography}{10}
\mciteSetBstSublistMode{n}
\mciteSetBstMaxWidthForm{subitem}{\alph{mcitesubitemcount})}
\mciteSetBstSublistLabelBeginEnd{\mcitemaxwidthsubitemform\space}
{\relax}{\relax}

\bibitem{Grozin:1992yq}
A.~G. Grozin, \ifthenelse{\boolean{articletitles}}{\emph{{Introduction to the
  heavy quark effective theory. Part 1}},
  }{}\href{http://arxiv.org/abs/hep-ph/9908366}{{\normalfont\ttfamily
  arXiv:hep-ph/9908366}}\relax
\mciteBstWouldAddEndPuncttrue
\mciteSetBstMidEndSepPunct{\mcitedefaultmidpunct}
{\mcitedefaultendpunct}{\mcitedefaultseppunct}\relax
\EndOfBibitem
\bibitem{Mannel:1996rg}
T.~Mannel, \ifthenelse{\boolean{articletitles}}{\emph{{Effective theory for
  heavy quarks}}, }{}\href{https://doi.org/10.1007/BFb0104296}{Lect.\ Notes
  Phys.\  \textbf{479} (1997) 387},
  \href{http://arxiv.org/abs/hep-ph/9606299}{{\normalfont\ttfamily
  arXiv:hep-ph/9606299}}\relax
\mciteBstWouldAddEndPuncttrue
\mciteSetBstMidEndSepPunct{\mcitedefaultmidpunct}
{\mcitedefaultendpunct}{\mcitedefaultseppunct}\relax
\EndOfBibitem
\bibitem{Ebert:2007nw}
D.~Ebert, R.~N. Faustov, and V.~O. Galkin,
  \ifthenelse{\boolean{articletitles}}{\emph{{Masses of excited heavy baryons
  in the relativistic quark-diquark model}},
  }{}\href{https://doi.org/10.1016/j.physletb.2007.11.037}{Phys.\ Lett.\
  \textbf{B659} (2008) 612},
  \href{http://arxiv.org/abs/0705.2957}{{\normalfont\ttfamily
  arXiv:0705.2957}}\relax
\mciteBstWouldAddEndPuncttrue
\mciteSetBstMidEndSepPunct{\mcitedefaultmidpunct}
{\mcitedefaultendpunct}{\mcitedefaultseppunct}\relax
\EndOfBibitem
\bibitem{Roberts:2007ni}
W.~Roberts and M.~Pervin, \ifthenelse{\boolean{articletitles}}{\emph{{Heavy
  baryons in a quark model}},
  }{}\href{https://doi.org/10.1142/S0217751X08041219}{Int.\ J.\ Mod.\ Phys.\
  \textbf{A23} (2008) 2817},
  \href{http://arxiv.org/abs/0711.2492}{{\normalfont\ttfamily
  arXiv:0711.2492}}\relax
\mciteBstWouldAddEndPuncttrue
\mciteSetBstMidEndSepPunct{\mcitedefaultmidpunct}
{\mcitedefaultendpunct}{\mcitedefaultseppunct}\relax
\EndOfBibitem
\bibitem{Garcilazo:2007eh}
H.~Garcilazo, J.~Vijande, and A.~Valcarce,
  \ifthenelse{\boolean{articletitles}}{\emph{{Faddeev study of heavy baryon
  spectroscopy}}, }{}\href{https://doi.org/10.1088/0954-3899/34/5/014}{J.\
  Phys.\  \textbf{G34} (2007) 961},
  \href{http://arxiv.org/abs/hep-ph/0703257}{{\normalfont\ttfamily
  arXiv:hep-ph/0703257}}\relax
\mciteBstWouldAddEndPuncttrue
\mciteSetBstMidEndSepPunct{\mcitedefaultmidpunct}
{\mcitedefaultendpunct}{\mcitedefaultseppunct}\relax
\EndOfBibitem
\bibitem{Migura:2006ep}
S.~Migura, D.~Merten, B.~Metsch, and H.-R. Petry,
  \ifthenelse{\boolean{articletitles}}{\emph{{Charmed baryons in a relativistic
  quark model}}, }{}\href{https://doi.org/10.1140/epja/i2006-10017-9}{Eur.\
  Phys.\ J.\  \textbf{A28} (2006) 41},
  \href{http://arxiv.org/abs/hep-ph/0602153}{{\normalfont\ttfamily
  arXiv:hep-ph/0602153}}\relax
\mciteBstWouldAddEndPuncttrue
\mciteSetBstMidEndSepPunct{\mcitedefaultmidpunct}
{\mcitedefaultendpunct}{\mcitedefaultseppunct}\relax
\EndOfBibitem
\bibitem{Ebert:2011kk}
D.~Ebert, R.~N. Faustov, and V.~O. Galkin,
  \ifthenelse{\boolean{articletitles}}{\emph{{Spectroscopy and Regge
  trajectories of heavy baryons in the relativistic quark-diquark picture}},
  }{}\href{https://doi.org/10.1103/PhysRevD.84.014025}{Phys.\ Rev.\
  \textbf{D84} (2011) 014025},
  \href{http://arxiv.org/abs/1105.0583}{{\normalfont\ttfamily
  arXiv:1105.0583}}\relax
\mciteBstWouldAddEndPuncttrue
\mciteSetBstMidEndSepPunct{\mcitedefaultmidpunct}
{\mcitedefaultendpunct}{\mcitedefaultseppunct}\relax
\EndOfBibitem
\bibitem{Valcarce:2008dr}
A.~Valcarce, H.~Garcilazo, and J.~Vijande,
  \ifthenelse{\boolean{articletitles}}{\emph{{Towards an understanding of heavy
  baryon spectroscopy}},
  }{}\href{https://doi.org/10.1140/epja/i2008-10616-4}{Eur.\ Phys.\ J.\
  \textbf{A37} (2008) 217},
  \href{http://arxiv.org/abs/0807.2973}{{\normalfont\ttfamily
  arXiv:0807.2973}}\relax
\mciteBstWouldAddEndPuncttrue
\mciteSetBstMidEndSepPunct{\mcitedefaultmidpunct}
{\mcitedefaultendpunct}{\mcitedefaultseppunct}\relax
\EndOfBibitem
\bibitem{Shah:2016nxi}
Z.~Shah, K.~Thakkar, A.~K. Rai, and P.~C. Vinodkumar,
  \ifthenelse{\boolean{articletitles}}{\emph{{Mass spectra and Regge
  trajectories of $\Lc$, $\Sigmares_\cquark^0$, $\Xicz$ and $\Omegac$
  baryons}}, }{}\href{https://doi.org/10.1088/1674-1137/40/12/123102}{Chin.\
  Phys.\  \textbf{C40} (2016) 123102},
  \href{http://arxiv.org/abs/1609.08464}{{\normalfont\ttfamily
  arXiv:1609.08464}}\relax
\mciteBstWouldAddEndPuncttrue
\mciteSetBstMidEndSepPunct{\mcitedefaultmidpunct}
{\mcitedefaultendpunct}{\mcitedefaultseppunct}\relax
\EndOfBibitem
\bibitem{Vijande:2012mk}
J.~Vijande, A.~Valcarce, T.~F. Carames, and H.~Garcilazo,
  \ifthenelse{\boolean{articletitles}}{\emph{{Heavy hadron spectroscopy: A
  quark model perspective}},
  }{}\href{https://doi.org/10.1142/S0218301313300117}{Int.\ J.\ Mod.\ Phys.\
  \textbf{E22} (2013) 1330011},
  \href{http://arxiv.org/abs/1212.4383}{{\normalfont\ttfamily
  arXiv:1212.4383}}\relax
\mciteBstWouldAddEndPuncttrue
\mciteSetBstMidEndSepPunct{\mcitedefaultmidpunct}
{\mcitedefaultendpunct}{\mcitedefaultseppunct}\relax
\EndOfBibitem
\bibitem{Yoshida:2015tia}
T.~Yoshida {\em et~al.}, \ifthenelse{\boolean{articletitles}}{\emph{{Spectrum
  of heavy baryons in the quark model}},
  }{}\href{https://doi.org/10.1103/PhysRevD.92.114029}{Phys.\ Rev.\
  \textbf{D92} (2015) 114029},
  \href{http://arxiv.org/abs/1510.01067}{{\normalfont\ttfamily
  arXiv:1510.01067}}\relax
\mciteBstWouldAddEndPuncttrue
\mciteSetBstMidEndSepPunct{\mcitedefaultmidpunct}
{\mcitedefaultendpunct}{\mcitedefaultseppunct}\relax
\EndOfBibitem
\bibitem{Chen:2015kpa}
H.-X. Chen {\em et~al.}, \ifthenelse{\boolean{articletitles}}{\emph{{P-wave
  charmed baryons from QCD sum rules}},
  }{}\href{https://doi.org/10.1103/PhysRevD.91.054034}{Phys.\ Rev.\
  \textbf{D91} (2015) 054034},
  \href{http://arxiv.org/abs/1502.01103}{{\normalfont\ttfamily
  arXiv:1502.01103}}\relax
\mciteBstWouldAddEndPuncttrue
\mciteSetBstMidEndSepPunct{\mcitedefaultmidpunct}
{\mcitedefaultendpunct}{\mcitedefaultseppunct}\relax
\EndOfBibitem
\bibitem{Chen:2016phw}
H.-X. Chen {\em et~al.}, \ifthenelse{\boolean{articletitles}}{\emph{{D-wave
  charmed and bottomed baryons from QCD sum rules}},
  }{}\href{https://doi.org/10.1103/PhysRevD.94.114016}{Phys.\ Rev.\
  \textbf{D94} (2016) 114016},
  \href{http://arxiv.org/abs/1611.02677}{{\normalfont\ttfamily
  arXiv:1611.02677}}\relax
\mciteBstWouldAddEndPuncttrue
\mciteSetBstMidEndSepPunct{\mcitedefaultmidpunct}
{\mcitedefaultendpunct}{\mcitedefaultseppunct}\relax
\EndOfBibitem
\bibitem{Padmanath:2013bla}
M.~Padmanath, R.~G. Edwards, N.~Mathur, and M.~Peardon,
  \ifthenelse{\boolean{articletitles}}{\emph{{Excited-state spectroscopy of
  singly, doubly and triply-charmed baryons from lattice QCD}}, }{} in {\em
  {Proceedings, 6th International Workshop on Charm Physics (Charm 2013):
  Manchester, UK, August 31-September 4, 2013}}, 2013,
  \href{http://arxiv.org/abs/1311.4806}{{\normalfont\ttfamily
  arXiv:1311.4806}}\relax
\mciteBstWouldAddEndPuncttrue
\mciteSetBstMidEndSepPunct{\mcitedefaultmidpunct}
{\mcitedefaultendpunct}{\mcitedefaultseppunct}\relax
\EndOfBibitem
\bibitem{LHCb-PAPER-2017-002}
LHCb collaboration, R.~Aaij {\em et~al.},
  \ifthenelse{\boolean{articletitles}}{\emph{{Observation of five new narrow
  $\Omegac$ states decaying to $\Xicp\Km$}},
  }{}\href{https://doi.org/10.1103/PhysRevLett.118.182001}{Phys.\ Rev.\ Lett.\
  \textbf{118} (2017) 182001},
  \href{http://arxiv.org/abs/1703.04639}{{\normalfont\ttfamily
  arXiv:1703.04639}}\relax
\mciteBstWouldAddEndPuncttrue
\mciteSetBstMidEndSepPunct{\mcitedefaultmidpunct}
{\mcitedefaultendpunct}{\mcitedefaultseppunct}\relax
\EndOfBibitem
\bibitem{Yelton:2017qxg}
Belle collaboration, J.~Yelton {\em et~al.},
  \ifthenelse{\boolean{articletitles}}{\emph{{Observation of excited
  $\POmega_\cquark$ charmed baryons in $e^+e^-$ collisions}},
  }{}\href{https://doi.org/10.1103/PhysRevD.97.051102}{Phys.\ Rev.\
  \textbf{D97} (2018) 051102},
  \href{http://arxiv.org/abs/1711.07927}{{\normalfont\ttfamily
  arXiv:1711.07927}}\relax
\mciteBstWouldAddEndPuncttrue
\mciteSetBstMidEndSepPunct{\mcitedefaultmidpunct}
{\mcitedefaultendpunct}{\mcitedefaultseppunct}\relax
\EndOfBibitem
\bibitem{Chiladze:1997ev}
G.~Chiladze and A.~F. Falk,
  \ifthenelse{\boolean{articletitles}}{\emph{{Phenomenology of new baryons with
  charm and strangeness}},
  }{}\href{https://doi.org/10.1103/PhysRevD.56.R6738}{Phys.\ Rev.\
  \textbf{D56} (1997) R6738},
  \href{http://arxiv.org/abs/hep-ph/9707507}{{\normalfont\ttfamily
  arXiv:hep-ph/9707507}}\relax
\mciteBstWouldAddEndPuncttrue
\mciteSetBstMidEndSepPunct{\mcitedefaultmidpunct}
{\mcitedefaultendpunct}{\mcitedefaultseppunct}\relax
\EndOfBibitem
\bibitem{Karliner:2017kfm}
M.~Karliner and J.~L. Rosner, \ifthenelse{\boolean{articletitles}}{\emph{{Very
  narrow excited $\POmega_\cquark$ baryons}},
  }{}\href{https://doi.org/10.1103/PhysRevD.95.114012}{Phys.\ Rev.\
  \textbf{D95} (2017) 114012},
  \href{http://arxiv.org/abs/1703.07774}{{\normalfont\ttfamily
  arXiv:1703.07774}}\relax
\mciteBstWouldAddEndPuncttrue
\mciteSetBstMidEndSepPunct{\mcitedefaultmidpunct}
{\mcitedefaultendpunct}{\mcitedefaultseppunct}\relax
\EndOfBibitem
\bibitem{LHCb-PAPER-2019-042}
LHCb collaboration, R.~Aaij {\em et~al.},
  \ifthenelse{\boolean{articletitles}}{\emph{{First observation of excited
  $\Omegares_b^-$ states}},
  }{}\href{https://doi.org/10.1103/PhysRevLett.124.082002}{Phys.\ Rev.\ Lett.\
  \textbf{124} (2020) 082002},
  \href{http://arxiv.org/abs/2001.00851}{{\normalfont\ttfamily
  arXiv:2001.00851}}\relax
\mciteBstWouldAddEndPuncttrue
\mciteSetBstMidEndSepPunct{\mcitedefaultmidpunct}
{\mcitedefaultendpunct}{\mcitedefaultseppunct}\relax
\EndOfBibitem
\bibitem{Aubert:2007eb}
BaBar collaboration, B.~Aubert {\em et~al.},
  \ifthenelse{\boolean{articletitles}}{\emph{{Study of
  $\Bbar\ensuremath{\rightarrow}{\ensuremath{\PXi}}_{\cquark} \Lcbar$ and
  $\Bbar\ensuremath{\rightarrow}\Lc \Lcbar\Kbar$ decays at BABAR}},
  }{}\href{https://doi.org/10.1103/PhysRevD.77.031101}{Phys.\ Rev.\
  \textbf{D77} (2008) 031101},
  \href{http://arxiv.org/abs/0710.5775}{{\normalfont\ttfamily
  arXiv:0710.5775}}\relax
\mciteBstWouldAddEndPuncttrue
\mciteSetBstMidEndSepPunct{\mcitedefaultmidpunct}
{\mcitedefaultendpunct}{\mcitedefaultseppunct}\relax
\EndOfBibitem
\bibitem{Aubert:2007dt}
BaBar collaboration, B.~Aubert {\em et~al.},
  \ifthenelse{\boolean{articletitles}}{\emph{{Study of excited charm-strange
  baryons with evidence for new baryons $\PXi_\cquark(3055)^+$ and
  $\PXi_\cquark(3123)^+$}},
  }{}\href{https://doi.org/10.1103/PhysRevD.77.012002}{Phys.\ Rev.\
  \textbf{D77} (2008) 012002},
  \href{http://arxiv.org/abs/0710.5763}{{\normalfont\ttfamily
  arXiv:0710.5763}}\relax
\mciteBstWouldAddEndPuncttrue
\mciteSetBstMidEndSepPunct{\mcitedefaultmidpunct}
{\mcitedefaultendpunct}{\mcitedefaultseppunct}\relax
\EndOfBibitem
\bibitem{Li:2017uvv}
Belle collaboration, Y.~B. Li {\em et~al.},
  \ifthenelse{\boolean{articletitles}}{\emph{{Observation of
  $\PXi_{\cquark}(2930)^0$ and updated measurement of $\Bm \to \Km \Lc \Lcbar$
  at Belle}}, }{}\href{https://doi.org/10.1140/epjc/s10052-018-5720-5}{Eur.\
  Phys.\ J.\  \textbf{C78} (2018) 252},
  \href{http://arxiv.org/abs/1712.03612}{{\normalfont\ttfamily
  arXiv:1712.03612}}\relax
\mciteBstWouldAddEndPuncttrue
\mciteSetBstMidEndSepPunct{\mcitedefaultmidpunct}
{\mcitedefaultendpunct}{\mcitedefaultseppunct}\relax
\EndOfBibitem
\bibitem{Li:2018fmq}
Belle collaboration, Y.~B. Li {\em et~al.},
  \ifthenelse{\boolean{articletitles}}{\emph{{Evidence of a structure in $\Kzb
  \Lc$ consistent with a charged $\PXi _\cquark(2930)^{+}$, and updated
  measurement of $\Bzb \rightarrow \Kzb \Lc \Lcbar$ at Belle}},
  }{}\href{https://doi.org/10.1140/epjc/s10052-018-6425-5}{Eur.\ Phys.\ J.\
  \textbf{C78} (2018) 928},
  \href{http://arxiv.org/abs/1806.09182}{{\normalfont\ttfamily
  arXiv:1806.09182}}\relax
\mciteBstWouldAddEndPuncttrue
\mciteSetBstMidEndSepPunct{\mcitedefaultmidpunct}
{\mcitedefaultendpunct}{\mcitedefaultseppunct}\relax
\EndOfBibitem
\bibitem{LHCb-DP-2008-001}
LHCb collaboration, A.~A. Alves~Jr.\ {\em et~al.},
  \ifthenelse{\boolean{articletitles}}{\emph{{The \lhcb detector at the LHC}},
  }{}\href{https://doi.org/10.1088/1748-0221/3/08/S08005}{JINST \textbf{3}
  (2008) S08005}\relax
\mciteBstWouldAddEndPuncttrue
\mciteSetBstMidEndSepPunct{\mcitedefaultmidpunct}
{\mcitedefaultendpunct}{\mcitedefaultseppunct}\relax
\EndOfBibitem
\bibitem{LHCb-DP-2014-002}
LHCb collaboration, R.~Aaij {\em et~al.},
  \ifthenelse{\boolean{articletitles}}{\emph{{LHCb detector performance}},
  }{}\href{https://doi.org/10.1142/S0217751X15300227}{Int.\ J.\ Mod.\ Phys.\
  \textbf{A30} (2015) 1530022},
  \href{http://arxiv.org/abs/1412.6352}{{\normalfont\ttfamily
  arXiv:1412.6352}}\relax
\mciteBstWouldAddEndPuncttrue
\mciteSetBstMidEndSepPunct{\mcitedefaultmidpunct}
{\mcitedefaultendpunct}{\mcitedefaultseppunct}\relax
\EndOfBibitem
\bibitem{LHCb-DP-2012-004}
R.~Aaij {\em et~al.}, \ifthenelse{\boolean{articletitles}}{\emph{{The \lhcb
  trigger and its performance in 2011}},
  }{}\href{https://doi.org/10.1088/1748-0221/8/04/P04022}{JINST \textbf{8}
  (2013) P04022}, \href{http://arxiv.org/abs/1211.3055}{{\normalfont\ttfamily
  arXiv:1211.3055}}\relax
\mciteBstWouldAddEndPuncttrue
\mciteSetBstMidEndSepPunct{\mcitedefaultmidpunct}
{\mcitedefaultendpunct}{\mcitedefaultseppunct}\relax
\EndOfBibitem
\bibitem{LHCb-DP-2016-001}
R.~Aaij {\em et~al.}, \ifthenelse{\boolean{articletitles}}{\emph{{Tesla: An
  application for real-time data analysis in High Energy Physics}},
  }{}\href{https://doi.org/10.1016/j.cpc.2016.07.022}{Comput.\ Phys.\ Commun.\
  \textbf{208} (2016) 35},
  \href{http://arxiv.org/abs/1604.05596}{{\normalfont\ttfamily
  arXiv:1604.05596}}\relax
\mciteBstWouldAddEndPuncttrue
\mciteSetBstMidEndSepPunct{\mcitedefaultmidpunct}
{\mcitedefaultendpunct}{\mcitedefaultseppunct}\relax
\EndOfBibitem
\bibitem{Sjostrand:2007gs}
T.~Sj\"{o}strand, S.~Mrenna, and P.~Skands,
  \ifthenelse{\boolean{articletitles}}{\emph{{A brief introduction to PYTHIA
  8.1}}, }{}\href{https://doi.org/10.1016/j.cpc.2008.01.036}{Comput.\ Phys.\
  Commun.\  \textbf{178} (2008) 852},
  \href{http://arxiv.org/abs/0710.3820}{{\normalfont\ttfamily
  arXiv:0710.3820}}\relax
\mciteBstWouldAddEndPuncttrue
\mciteSetBstMidEndSepPunct{\mcitedefaultmidpunct}
{\mcitedefaultendpunct}{\mcitedefaultseppunct}\relax
\EndOfBibitem
\bibitem{LHCb-PROC-2010-056}
I.~Belyaev {\em et~al.}, \ifthenelse{\boolean{articletitles}}{\emph{{Handling
  of the generation of primary events in Gauss, the LHCb simulation
  framework}}, }{}\href{https://doi.org/10.1088/1742-6596/331/3/032047}{J.\
  Phys.\ Conf.\ Ser.\  \textbf{331} (2011) 032047}\relax
\mciteBstWouldAddEndPuncttrue
\mciteSetBstMidEndSepPunct{\mcitedefaultmidpunct}
{\mcitedefaultendpunct}{\mcitedefaultseppunct}\relax
\EndOfBibitem
\bibitem{Lange:2001uf}
D.~J. Lange, \ifthenelse{\boolean{articletitles}}{\emph{{The EvtGen particle
  decay simulation package}},
  }{}\href{https://doi.org/10.1016/S0168-9002(01)00089-4}{Nucl.\ Instrum.\
  Meth.\  \textbf{A462} (2001) 152}\relax
\mciteBstWouldAddEndPuncttrue
\mciteSetBstMidEndSepPunct{\mcitedefaultmidpunct}
{\mcitedefaultendpunct}{\mcitedefaultseppunct}\relax
\EndOfBibitem
\bibitem{Allison:2006ve}
Geant4 collaboration, J.~Allison {\em et~al.},
  \ifthenelse{\boolean{articletitles}}{\emph{{Geant4 developments and
  applications}}, }{}\href{https://doi.org/10.1109/TNS.2006.869826}{IEEE
  Trans.\ Nucl.\ Sci.\  \textbf{53} (2006) 270}\relax
\mciteBstWouldAddEndPuncttrue
\mciteSetBstMidEndSepPunct{\mcitedefaultmidpunct}
{\mcitedefaultendpunct}{\mcitedefaultseppunct}\relax
\EndOfBibitem
\bibitem{LHCb-PROC-2011-006}
M.~Clemencic {\em et~al.}, \ifthenelse{\boolean{articletitles}}{\emph{{The
  \lhcb simulation application, Gauss: Design, evolution and experience}},
  }{}\href{https://doi.org/10.1088/1742-6596/331/3/032023}{J.\ Phys.\ Conf.\
  Ser.\  \textbf{331} (2011) 032023}\relax
\mciteBstWouldAddEndPuncttrue
\mciteSetBstMidEndSepPunct{\mcitedefaultmidpunct}
{\mcitedefaultendpunct}{\mcitedefaultseppunct}\relax
\EndOfBibitem
\bibitem{Breiman}
L.~Breiman, J.~H. Friedman, R.~A. Olshen, and C.~J. Stone, {\em Classification
  and regression trees}, Wadsworth international group, Belmont, California,
  USA, 1984\relax
\mciteBstWouldAddEndPuncttrue
\mciteSetBstMidEndSepPunct{\mcitedefaultmidpunct}
{\mcitedefaultendpunct}{\mcitedefaultseppunct}\relax
\EndOfBibitem
\bibitem{AdaBoost}
Y.~Freund and R.~E. Schapire, \ifthenelse{\boolean{articletitles}}{\emph{A
  decision-theoretic generalization of on-line learning and an application to
  boosting}, }{}\href{https://doi.org/10.1006/jcss.1997.1504}{J.\ Comput.\
  Syst.\ Sci.\  \textbf{55} (1997) 119}\relax
\mciteBstWouldAddEndPuncttrue
\mciteSetBstMidEndSepPunct{\mcitedefaultmidpunct}
{\mcitedefaultendpunct}{\mcitedefaultseppunct}\relax
\EndOfBibitem
\bibitem{Hocker:2007ht}
H.~Voss, A.~Hoecker, J.~Stelzer, and F.~Tegenfeldt,
  \ifthenelse{\boolean{articletitles}}{\emph{{TMVA - the Toolkit for
  Multivariate Data Analysis with ROOT}},
  }{}\href{https://doi.org/10.22323/1.050.0040}{PoS \textbf{ACAT} (2007)
  040}\relax
\mciteBstWouldAddEndPuncttrue
\mciteSetBstMidEndSepPunct{\mcitedefaultmidpunct}
{\mcitedefaultendpunct}{\mcitedefaultseppunct}\relax
\EndOfBibitem
\bibitem{TMVA4}
A.~Hoecker {\em et~al.}, \ifthenelse{\boolean{articletitles}}{\emph{{TMVA 4 -
  Toolkit for Multivariate Data Analysis with ROOT. Users Guide.}},
  }{}\href{http://arxiv.org/abs/physics/0703039}{{\normalfont\ttfamily
  arXiv:physics/0703039}}\relax
\mciteBstWouldAddEndPuncttrue
\mciteSetBstMidEndSepPunct{\mcitedefaultmidpunct}
{\mcitedefaultendpunct}{\mcitedefaultseppunct}\relax
\EndOfBibitem
\bibitem{Pivk:2004ty}
M.~Pivk and F.~R. Le~Diberder,
  \ifthenelse{\boolean{articletitles}}{\emph{{sPlot: A statistical tool to
  unfold data distributions}},
  }{}\href{https://doi.org/10.1016/j.nima.2005.08.106}{Nucl.\ Instrum.\ Meth.\
  \textbf{A555} (2005) 356},
  \href{http://arxiv.org/abs/physics/0402083}{{\normalfont\ttfamily
  arXiv:physics/0402083}}\relax
\mciteBstWouldAddEndPuncttrue
\mciteSetBstMidEndSepPunct{\mcitedefaultmidpunct}
{\mcitedefaultendpunct}{\mcitedefaultseppunct}\relax
\EndOfBibitem
\bibitem{PDG2019}
Particle Data Group, M.~Tanabashi {\em et~al.},
  \ifthenelse{\boolean{articletitles}}{\emph{{\href{http://pdg.lbl.gov/}{Review
  of particle physics}}},
  }{}\href{https://doi.org/10.1103/PhysRevD.98.030001}{Phys.\ Rev.\
  \textbf{D98} (2018) 030001}, and {\href{http://pdglive.lbl.gov/}{2019
  update}}\relax
\mciteBstWouldAddEndPuncttrue
\mciteSetBstMidEndSepPunct{\mcitedefaultmidpunct}
{\mcitedefaultendpunct}{\mcitedefaultseppunct}\relax
\EndOfBibitem
\bibitem{Punzi:2003bu}
G.~Punzi, \ifthenelse{\boolean{articletitles}}{\emph{{Sensitivity of searches
  for new signals and its optimization}}, }{}eConf \textbf{C030908} (2003)
  MODT002, \href{http://arxiv.org/abs/physics/0308063}{{\normalfont\ttfamily
  arXiv:physics/0308063}}\relax
\mciteBstWouldAddEndPuncttrue
\mciteSetBstMidEndSepPunct{\mcitedefaultmidpunct}
{\mcitedefaultendpunct}{\mcitedefaultseppunct}\relax
\EndOfBibitem
\bibitem{Cowan:2016tnm}
G.~A. Cowan, D.~C. Craik, and M.~D. Needham,
  \ifthenelse{\boolean{articletitles}}{\emph{{RapidSim: An application for the
  fast simulation of heavy-quark hadron decays}},
  }{}\href{https://doi.org/10.1016/j.cpc.2017.01.029}{Comput.\ Phys.\ Commun.\
  \textbf{214} (2017) 239},
  \href{http://arxiv.org/abs/1612.07489}{{\normalfont\ttfamily
  arXiv:1612.07489}}\relax
\mciteBstWouldAddEndPuncttrue
\mciteSetBstMidEndSepPunct{\mcitedefaultmidpunct}
{\mcitedefaultendpunct}{\mcitedefaultseppunct}\relax
\EndOfBibitem
\bibitem{Blatt:1952ije}
J.~M. Blatt and V.~F. Weisskopf, {\em {Theoretical nuclear physics}},
  \href{https://doi.org/10.1007/978-1-4612-9959-2}{ Springer, New York,
  1952}\relax
\mciteBstWouldAddEndPuncttrue
\mciteSetBstMidEndSepPunct{\mcitedefaultmidpunct}
{\mcitedefaultendpunct}{\mcitedefaultseppunct}\relax
\EndOfBibitem
\bibitem{LHCb-PAPER-2013-011}
LHCb collaboration, R.~Aaij {\em et~al.},
  \ifthenelse{\boolean{articletitles}}{\emph{{Precision measurement of \D meson
  mass differences}}, }{}\href{https://doi.org/10.1007/JHEP06(2013)065}{JHEP
  \textbf{06} (2013) 065},
  \href{http://arxiv.org/abs/1304.6865}{{\normalfont\ttfamily
  arXiv:1304.6865}}\relax
\mciteBstWouldAddEndPuncttrue
\mciteSetBstMidEndSepPunct{\mcitedefaultmidpunct}
{\mcitedefaultendpunct}{\mcitedefaultseppunct}\relax
\EndOfBibitem
\bibitem{LHCb-PAPER-2010-001}
LHCb collaboration, R.~Aaij {\em et~al.},
  \ifthenelse{\boolean{articletitles}}{\emph{{Prompt \KS production in
  \proton\proton collisions at $\sqs=$0.9\tev}},
  }{}\href{https://doi.org/10.1016/j.physletb.2010.08.055}{Phys.\ Lett.\
  \textbf{B693} (2010) 69},
  \href{http://arxiv.org/abs/1008.3105}{{\normalfont\ttfamily
  arXiv:1008.3105}}\relax
\mciteBstWouldAddEndPuncttrue
\mciteSetBstMidEndSepPunct{\mcitedefaultmidpunct}
{\mcitedefaultendpunct}{\mcitedefaultseppunct}\relax
\EndOfBibitem
\bibitem{LHCb-PAPER-2014-048}
LHCb collaboration, R.~Aaij {\em et~al.},
  \ifthenelse{\boolean{articletitles}}{\emph{{Precision measurement of the mass
  and lifetime of the \Xibm baryon}},
  }{}\href{https://doi.org/10.1103/PhysRevLett.113.242002}{Phys.\ Rev.\ Lett.\
  \textbf{113} (2014) 242002},
  \href{http://arxiv.org/abs/1409.8568}{{\normalfont\ttfamily
  arXiv:1409.8568}}\relax
\mciteBstWouldAddEndPuncttrue
\mciteSetBstMidEndSepPunct{\mcitedefaultmidpunct}
{\mcitedefaultendpunct}{\mcitedefaultseppunct}\relax
\EndOfBibitem
\bibitem{Yelton:2016fqw}
Belle collaboration, J.~Yelton {\em et~al.},
  \ifthenelse{\boolean{articletitles}}{\emph{{Study of excited $\Xic$ states
  decaying into $\Xicz$ and $\Xicp$ baryons}},
  }{}\href{https://doi.org/10.1103/PhysRevD.94.052011}{Phys.\ Rev.\
  \textbf{D94} (2016) 052011},
  \href{http://arxiv.org/abs/1607.07123}{{\normalfont\ttfamily
  arXiv:1607.07123}}\relax
\mciteBstWouldAddEndPuncttrue
\mciteSetBstMidEndSepPunct{\mcitedefaultmidpunct}
{\mcitedefaultendpunct}{\mcitedefaultseppunct}\relax
\EndOfBibitem
\bibitem{Lesiak:2008wz}
Belle collaboration, T.~Lesiak {\em et~al.},
  \ifthenelse{\boolean{articletitles}}{\emph{{Measurement of masses of the
  $\Xic(2645)$ and $\Xic(2815)$ baryons and observation of $\Xic(2980)\to
  \Xic(2645)\pi$}},
  }{}\href{https://doi.org/10.1016/j.physletb.2008.05.055}{Phys.\ Lett.\
  \textbf{B665} (2008) 9},
  \href{http://arxiv.org/abs/0802.3968}{{\normalfont\ttfamily
  arXiv:0802.3968}}\relax
\mciteBstWouldAddEndPuncttrue
\mciteSetBstMidEndSepPunct{\mcitedefaultmidpunct}
{\mcitedefaultendpunct}{\mcitedefaultseppunct}\relax
\EndOfBibitem
\bibitem{GellMann:1962xb}
M.~Gell-Mann, \ifthenelse{\boolean{articletitles}}{\emph{{Symmetries of baryons
  and mesons}}, }{}\href{https://doi.org/10.1103/PhysRev.125.1067}{Phys.\ Rev.\
   \textbf{125} (1962) 1067}\relax
\mciteBstWouldAddEndPuncttrue
\mciteSetBstMidEndSepPunct{\mcitedefaultmidpunct}
{\mcitedefaultendpunct}{\mcitedefaultseppunct}\relax
\EndOfBibitem
\bibitem{Okubo:1961jc}
S.~Okubo, \ifthenelse{\boolean{articletitles}}{\emph{{Note on unitary symmetry
  in strong interactions}}, }{}\href{https://doi.org/10.1143/PTP.27.949}{Prog.\
  Theor.\ Phys.\  \textbf{27} (1962) 949}\relax
\mciteBstWouldAddEndPuncttrue
\mciteSetBstMidEndSepPunct{\mcitedefaultmidpunct}
{\mcitedefaultendpunct}{\mcitedefaultseppunct}\relax
\EndOfBibitem
\end{mcitethebibliography}
 }
 \ifthenelse{\boolean{wordcount}}{}{

\clearpage 

\newpage
\centerline
{\large\bf LHCb collaboration}
\begin
{flushleft}
\small
R.~Aaij$^{31}$,
C.~Abell{\'a}n~Beteta$^{49}$,
T.~Ackernley$^{59}$,
B.~Adeva$^{45}$,
M.~Adinolfi$^{53}$,
H.~Afsharnia$^{9}$,
C.A.~Aidala$^{81}$,
S.~Aiola$^{25}$,
Z.~Ajaltouni$^{9}$,
S.~Akar$^{66}$
% ...

{\footnotesize \it

% ...
$ ^{9}$Universit{\'e} Clermont Auvergne, CNRS/IN2P3, LPC, Clermont-Ferrand, France\\
% ...
$ ^{25}$INFN Sezione di Milano, Milano, Italy\\
% ...
$ ^{31}$Nikhef National Institute for Subatomic Physics, Amsterdam, Netherlands\\
% ...
$ ^{45}$Instituto Galego de F{\'\i}sica de Altas Enerx{\'\i}as (IGFAE), Universidade de Santiago de Compostela, Santiago de Compostela, Spain\\
% ...
$ ^{49}$Physik-Institut, Universit{\"a}t Z{\"u}rich, Z{\"u}rich, Switzerland\\
% ...
$ ^{53}$H.H. Wills Physics Laboratory, University of Bristol, Bristol, United Kingdom\\
% ...
$ ^{59}$Oliver Lodge Laboratory, University of Liverpool, Liverpool, United Kingdom\\
% ...
$ ^{66}$Los Alamos National Laboratory (LANL), Los Alamos, United States\\
$ ^{67}$Syracuse University, Syracuse, NY, United States\\
% ...
$ ^{81}$University of Michigan, Ann Arbor, United States, associated to $^{67}$\\
\bigskip
% ...
\medskip
}
\end{flushleft}

}

\end{document}